# Adaptive Quantization Matrices for HD and UHD Display Resolutions in Scalable HEVC

Lee Prangnell and Victor Sanchez

*Department of Computer Science, University of Warwick, England, UK*

**Abstract**: HEVC contains an option to enable custom quantization matrices (QMs), which are designed based on the Human Visual System (HVS) and a 2D Contrast Sensitivity Function (CSF). Visual Display Units (VDUs), capable of displaying video data at High Definition (HD) and Ultra HD (UHD) display resolutions, are frequently utilized on a global scale. Video compression artifacts that are present due to high levels of quantization, which are typically inconspicuous in low display resolution environments, are clearly visible on HD and UHD video data and VDUs. The default HVS-CSF QM technique in HEVC does not take into account the video data resolution, nor does it take into consideration the associated VDU's display resolution to determine the appropriate levels of quantization required to reduce unwanted video compression artifacts. Based on this fact, we propose a novel, adaptive quantization matrix technique for the HEVC standard, including Scalable HEVC (SHVC). Our technique, which is based on a refinement of the current HVS-CSF QM approach in HEVC, takes into consideration the display resolution of the target VDU for the purpose of minimizing video compression artifacts. In SHVC SHM 9.0, and compared with anchors, the proposed technique yields important quality and coding improvements for the Random Access configuration, with a maximum of 56.5% luma BD-Rate reductions in the enhancement layer. Furthermore, compared with the default QMs and the Sony QMs, our method yields encoding time reductions of 0.75% and 1.19%, respectively.

**1.0 Introduction**

When the scaling list option is enabled in HEVC and its standardized extensions, including SHVC, transform coefficients are quantized according to the weighting values in the QMs. More specifically, after the linear transformation of the residual values, by a finite precision approximation of the Discrete Cosine Transform (DCT), luma and chroma transform coefficients in a Transform Block (TB) are individually quantized according to the integer weighting values that are present in the intra and inter QMs. The default intra and inter QMs in HEVC are based on a 2D HVS-CSF model [1-4]. The integer values in the QMs correspond to the quantization weighting of low, medium and high frequency transform coefficients in a TB; therefore, these QMs possess the capacity to control the quantization step size. A TB contains DC and AC transform coefficients, where the DC transform coefficient is the lowest frequency component and where the AC coefficients correspond to low, medium and high frequency components [3]. Because low frequency transform coefficients are more important for video signal reconstruction, the default QMs in HEVC apply coarser quantization to medium and high frequency AC transform coefficients. Originally designed for the JPEG image compression standard for still image coding, the QM in [1] is presently employed as the default intra QM in HEVC. This intra QM is derived from a Frequency Weighting Matrix (FWM). The inter QM in HEVC is derived from the intra QM using a linear model [5].

Alternative QM methods have been proposed to improve upon the default intra and inter QMs in HEVC. To the best of our knowledge, no previous research has been undertaken in terms of designing a QM technique where the QMs adapt to the display resolution of the target VDU. Therefore, adapting QMs to VDUs is a novel concept. The QM method in [5] involves adjustments to the parameter selection of the HVS-CSF QM technique in [1]. This refinement produces a modified FWM, from which the intra and inter QMs are derived. Although these new parameter insertions may potentially produce coding efficiency improvements, this technique does not take into account the target VDU's display resolution in terms of the quantization of low, medium and high frequency transform coefficients. In [6], the authors propose a novel intra QM method that modifies the weighting values in the QM by employing a normalized exponent variable. Accordingly, the values in the FWM, which correspond to medium and high frequency transform coefficients, are modified to decrease the corresponding quantization levels. This results in a quality improvement of the finer details in the images. Similar to the method in [5], this technique does not take into account the target VDU's display resolution. Moreover, the exponent variable utilized to modify the FWM values is arbitrary.

Based on the display resolution of the target VDU, we propose a novel refinement of the HVS-CSF QM method presented in [1]. Video compression artifacts are much more visible at high display resolutions, such as HD and UHD, compared with low display resolution [7, 8]. This is true for raw video sequences specifically designed for HD and UHD resolutions and also for those designed for Standard Definition (SD) resolutions that are subsequently coded, decoded and deployed to HD and UHD VDUs. Our AQM technique provides a solution to this problem. The proposed intra and inter AQMs are integrated into SHVC to create a multilayered bit-stream that contains a Base Layer (BL) and Enhancement Layers (ELs). Each layer in this multilayered bit-stream is aimed at a specific display resolution of a VDU. More specifically, lower levels of quantization are applied to the ELs, which are then decoded and deployed to large VDU display resolutions (for example, 4K and 8K UHD). Conversely, a higher level of quantization is applied to the BL, which is then decoded and deployed to smaller VDU display resolutions (for example, HD and SD). The main objective of the proposed AQM technique is to decrease the visibility of any compression artifacts that are present, due to quantization, in the decoded layers deployed to large display resolution VDUs.

The rest of the paper is organized as follows. Section 2 provides an overview of the default QMs in HEVC. Section 3 includes detailed expositions of the proposed AQM technique. We briefly explain the utilization of the AQMs in SHVC in Section 4. Section 5 presents the evaluation results of the AQM technique. Finally, Section 6 concludes this paper.

## 2.0 Default QMs in HEVC

The HVS-CSF based QM technique in [1] is employed as the default intra QM in HEVC because of its advantages for frequency dependent scaling. The HVS-CSF based 8×8 intra QM, and the 8×8 inter QM that is derived from the intra QM, have been shown to be effective QM solutions in HEVC. The default QMs in HEVC allow low frequency AC transform coefficients to be quantized with a finer quantization step size in 8×8 TBs [9]. Although the HEVC standard supports up to 32×32 TBs, default 16×16 and 32×32 QMs are not present in the HEVC design; they are obtained from upsampling and replication of the 8×8 QMs. More specifically, to create a 16×16 QM, each entry in an 8×8 QM is upsampled and replicated into a 2×2 region, while each entry in an 8×8 QM is upsampled and replicated into a 4×4 region to create a 32×32 QM [9]. This QM replication process ensures that transform coefficients, in 16×16 and 32×32 TBs, are quantized appropriately according to their frequency content (see Fig. 1); note that we exploit this 8×8 QM upsampling and replication process for the proposed AQM technique. Because HEVC may employ up to a total of 20 QMs, this 8×8 QM upsampling and replication process is designed to minimize computational complexity with respect to the memory requirements needed to store the QMs [9].

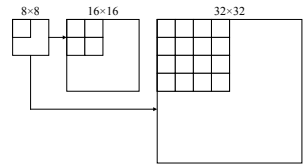

Fig. 1. To create a 16×16 QM, each entry in an 8x8 QM is upsampled and replicated into a 2×2 region, while each entry in an 8×8 QM is upsampled and replicated into a 4×4 region to create a 32×32 QM.

As previously mentioned, the default 8×8 intra QM in HEVC is derived from a HVS-CSF based approach [1, 2]. In this technique, the HVS is modeled as a nonlinear point transformation followed by the Modulation Transfer Function (MTF) [10]. A CSF-based MTF is first defined as follows:

$$H(f) = a(b + cf)\exp(-c(f)^d) \qquad (1)$$

where $f$ is the radial frequency in cycles per degree of the visual angle subtended, and $a$, $b$, $c$ and $d$ are constants.

Based on Daly's 2D HVS-CSF approach in [2], the MTF is computed with the modified constant values $a$=2.2, $b$=0.192, $c$=0.114 and $d$=1.1 [1]. The MTF is then used to produce a 2D FWM, $H(u,v)$, comprising floating point values from which the QM integer threshold values are derived. $H(u,v)$ is computed in (2):

$$H(u,v) = \begin{cases} 2.2(0.192 + 0.114 f'(u,v)) \exp(-(0.114 f'(u,v))^{1.1}) & \text{if } f'(u,v) > f_{max,} \\ 1.0 & \text{otherwise,} \end{cases} \quad (2)$$

where $u$ and $v$, in $H(u,v)$, represent the horizontal and vertical floating point values, $f'(u,v)$ is the normalized radial spatial frequency in cycles per degree and $f_{max}$ denotes the frequency of 8 cycles per degree (i.e., the exponential peak).

In order to account for the fluctuations in the MTF as a function of viewing angle $\theta$, the normalized radial spatial frequency, $f'(u,v)$, is defined using angular dependent function $S(\theta(u,v))$. Both $f'(u,v)$ and $S(\theta(u,v))$ are quantified in (3)-(6).

$$f'(u,v) = \frac{f(u,v)}{S(\theta(u,v))} \quad (3)$$

$$f(u,v) = \frac{\pi}{180 \sin^{-1}(1/\sqrt{1+dis^2})} \times \sqrt{f(u)^2 + f(v)^2} \quad (4)$$

$$S(\theta(u,v)) = \frac{1-s}{2} \cos(4\theta(u,v)) + \frac{1+s}{2} \quad (5)$$

$$\theta(u,v) = \arctan\left(\frac{f(u)}{f(v)}\right) \quad (6)$$

where *dis* represents the perceptual viewing distance of 512mm and $s$ is the symmetry parameter with a value of 0.7 [10]. Parameter $s$ ensures that the floating point values in $H(u,v)$ are symmetric. As $s$ decreases, $S(\theta(u,v))$ decreases at approximately 45°; this, in turn, increases $f'(u,v)$ and decreases $H(u,v)$. The discrete horizontal and vertical frequencies are computed in (7):

$$f(u) = \frac{u-1}{\Delta \times 2N}, \quad \text{for } u = 1, 2..., N;$$

$$f(v) = \frac{v-1}{\Delta \times 2N}, \quad \text{for } v = 1, 2..., N; \quad (7)$$

where $\Delta$ denotes the dot pitch value of 0.25mm (approximately 100 DPI) and $N$ is the number of horizontal and vertical radial spatial frequencies. The resulting 8×8 $H(u,v)$ matrix is shown in (8).

$$H(u,v) = \begin{pmatrix} 1.0000 & 1.0000 & 1.0000 & 1.0000 & 0.9599 & 0.8746 & 0.7684 & 0.6571 \\ 1.0000 & 1.0000 & 1.0000 & 1.0000 & 0.9283 & 0.8404 & 0.7371 & 0.6306 \\ 1.0000 & 1.0000 & 0.9571 & 0.8898 & 0.8192 & 0.7371 & 0.6471 & 0.5558 \\ 1.0000 & 1.0000 & 0.8898 & 0.7617 & 0.6669 & 0.5912 & 0.5196 & 0.4495 \\ 0.9599 & 0.9283 & 0.8192 & 0.6669 & 0.5419 & 0.4564 & 0.3930 & 0.3393 \\ 0.8746 & 0.8404 & 0.7371 & 0.5912 & 0.4564 & 0.3598 & 0.2948 & 0.2480 \\ 0.7684 & 0.7371 & 0.6471 & 0.5196 & 0.3930 & 0.2948 & 0.2278 & 0.1828 \\ 0.6571 & 0.6306 & 0.5558 & 0.4495 & 0.3393 & 0.2480 & 0.1828 & 0.1391 \end{pmatrix} \in [0,1] \quad (8)$$

The normalized values in $H(u,v)$ highlight the visually perceptual importance of transform coefficients in the frequency domain. These normalized values are then rounded to integer values to create an 8×8 intra QM, $QM_{intra}$ by utilizing a scaling value of 16 [11, 1]. The resulting default 8×8 $QM_{intra}$ is quantified in (9).

$$QM_{intra} = \left\lfloor \frac{16}{H(u,v)} \right\rfloor = \begin{pmatrix} 16 & 16 & 16 & 16 & 17 & 18 & 21 & 24 \\ 16 & 16 & 16 & 16 & 17 & 19 & 22 & 25 \\ 16 & 16 & 17 & 18 & 20 & 22 & 25 & 29 \\ 16 & 16 & 18 & 21 & 24 & 27 & 31 & 36 \\ 17 & 17 & 20 & 24 & 30 & 35 & 41 & 47 \\ 18 & 19 & 22 & 27 & 35 & 44 & 54 & 65 \\ 21 & 22 & 25 & 31 & 41 & 54 & 70 & 88 \\ 24 & 25 & 29 & 36 & 47 & 65 & 88 & 115 \end{pmatrix} \quad (9)$$

The default 8×8 inter QM in HEVC, $QM_{inter}$, is generated using a simple linear model, as specified in [5] and [12]. The resulting default 8×8 $QM_{inter}$ is shown in (10).

$$QM_{inter} = \begin{pmatrix} 16 & 16 & 16 & 16 & 17 & 18 & 20 & 24 \\ 16 & 16 & 16 & 17 & 18 & 20 & 24 & 25 \\ 16 & 16 & 17 & 18 & 20 & 24 & 25 & 28 \\ 16 & 17 & 18 & 20 & 24 & 25 & 28 & 33 \\ 17 & 18 & 20 & 24 & 25 & 28 & 33 & 41 \\ 18 & 20 & 24 & 25 & 28 & 33 & 41 & 54 \\ 20 & 24 & 25 & 28 & 33 & 41 & 54 & 71 \\ 24 & 25 & 28 & 33 & 41 & 54 & 71 & 91 \end{pmatrix} \quad (10)$$

**3.0 Proposed AQM Technique**

The proposed AQM method refines the intra QM technique derived from the HVS-CSF approach previously described [1]. This is achieved by applying a parameter that adapts the 2D FWM, $H(u,v)$, to the display resolution of the target VDU. The intra and inter AQMs are derived from the resulting modified 2D FWM, in which the values are calculated based on the target VDU's display resolution and also the visually perceptual importance of transform coefficients. In this work, we focus specifically on integrating our technique into SHVC in order to produce a single multilayered bit-stream, in which each layer is coded to attain the highest possible visual quality for the display resolution of the target VDU. Based on the observation that video compression artifacts are much more visible on high display resolution VDUs [7, 8], higher levels of quantization are applied to the BL and, conversely, lower levels of quantization are applied to the ELs. The BL is, therefore, aimed at smaller display resolution VDUs (for example, HD and SD), while the ELs are aimed at larger display resolution VDUs (for example, HD and UHD). At the TB level, similar to the default HEVC QMs, the proposed AQM technique still applies different levels of quantization to transform coefficients according to the frequency they represent. However, these quantization levels are now adapted to the display resolution of the target VDU.

Our technique is based on parameter $A_{i,j}$, which is applied to each element of $H(u,v)$ located at position $(i,j)$, denoted as $H_{i,j}$. $A_{i,j}$ allows $H_{i,j}$ to be modified according to the TB size and also according to the VDU's display resolution in order to produce the adaptive 2D FWM $H'(u,v)$. The element of $H'(u,v)$ located at position $(i,j)$, denoted as $H'_{i,j}$, is computed in (11).

$$H'_{i,j} = H_{i,j}^{A_{i,j}} \quad (11)$$

Note that the computations derived from equations (3) to (7) still apply to $H'(u,v)$. Parameter $A_{i,j}$ is calculated by a negative exponential function that uses as input the Euclidean distance between two coefficient locations in a TB and also a display resolution value. An exponential function is employed to modify $H(u,v)$ in order to produce, in the FWM, larger values for low frequency transform coefficients, and also to produce smaller values for high frequency transform coefficients. Consequently, this results in integer weighting values in the intra and inter AQMs that vary according to the frequency associated with the transform coefficients in a TB. The derived AQMs are then used in the quantization process. Parameter $A_{i,j}$ is computed as a function of two parameters in (12):

$$A_{i,j}\left(d_{i,j}, w\right) = e^{-\left(\frac{d_{i,j}}{w}\right)} \in [0,1] \tag{12}$$

where $d_{i,j}$ is the normalized Euclidean distance between the DC transform coefficient and the current coefficient located at position $(i,j)$ in an 8×8 TB, and $w$ is the display resolution parameter. Note that $w$ is a key parameter that controls the values in $H'(u,v)$ according to the display resolution of the target VDU. Euclidean distance $d_{i,j}$ is computed in (13):

$$d_{i,j} = \sqrt{\frac{(i_1 - i_2)^2 + (j_1 - j_2)^2}{(i_1 - i_{max})^2 + (j_1 - j_{max})^2}} \in [0,1] \tag{13}$$

where $(i_1, j_1)$, $(i_2, j_2)$, $(i_{max}, j_{max})$ represent the position of the floating point values in FWM $H(u,v)$ associated with the DC coefficient, the current coefficient and the farthest AC coefficient, respectively. The floating point value associated with the DC transform coefficient is located at position $(i = 0, j = 0)$, while the farthest AC coefficient is located at position $(i = 7, j = 7)$. Each $A_{i,j}$ value decreases as the display resolution parameter $w$ decreases. The display resolution parameter $w$ is quantified in (14):

$$w = h_t^{-p} \in (0,1] \tag{14}$$

where $h_t$ is the VDU's theoretical maximum hypotenuse value, in pixels, and where $p$ is the normalized hypotenuse value, in pixels. Value $p$ is computed in (15) and $h_t$ is calculated in (17):

$$p = \frac{h_a}{h_t} \in (0,1] \tag{15}$$

where $h_a$ is the VDU's actual hypotenuse value in the pixel domain and $h_t$ is the maximum possible hypotenuse value, which is used to normalize $h_a$. These values are computed as follows:

$$h_a = \sqrt{x^2 + y^2} \tag{16}$$

$$h_t = \sqrt{x_{max}^2 + y_{max}^2} \tag{17}$$

where $x$ and $y$ represent the horizontal and vertical dimensions of the target VDU, respectively (see Fig. 2), and where $x_{max}$ and $y_{max}$ represent, respectively, the maximum possible horizontal and vertical dimensions of the target VDU. Note that display resolution parameter $w$ produces a suitable normalized distribution of values based on a theoretical maximum display resolution. More specifically, value $w$ rapidly decreases as $p$ increases (i.e., as the target VDU's display resolution increases), which in turn results in higher values in adaptive FWM $H'(u,v)$, lower weighting values in the AQMs and, consequently, lower quantization levels.

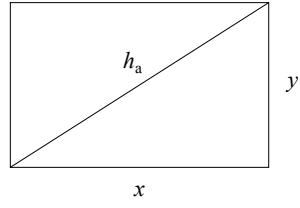

Fig. 2. An illustration of the hypotenuse $h_a$ and also cathetus 1 and 2, in pixels, of a VDU's display resolution.

Because the default 8×8 intra QM in [1] is designed for the JPEG standard, we base the theoretical maximum pixel values, $x_{max}$ and $y_{max}$, on the maximum possible image size, in pixels, permitted in the JPEG standard. These pixel values are as follows: 65535×65535 [14]. Therefore, $x_{max}$=65535 and $y_{max}$=65535.

The 8×8 matrices in (18) and (19) are the 2D FWM $H'(u,v)$ and the intra AQM $AQM_{intra}$, respectively, as computed using (11)-(17) for a target VDU of 3840×2160 pixels (4K).

$$H'(u,v) = \begin{pmatrix} 1.0000 & 1.0000 & 1.0000 & 1.0000 & 0.9798 & 0.9454 & 0.9114 & 0.8832 \\ 1.0000 & 1.0000 & 1.0000 & 1.0000 & 0.9643 & 0.9309 & 0.8996 & 0.8739 \\ 1.0000 & 1.0000 & 0.9736 & 0.9396 & 0.9125 & 0.8873 & 0.8652 & 0.8475 \\ 1.0000 & 1.0000 & 0.9396 & 0.8780 & 0.8439 & 0.8265 & 0.8156 & 0.8085 \\ 0.9798 & 0.9643 & 0.9125 & 0.8439 & 0.7953 & 0.7730 & 0.7662 & 0.7666 \\ 0.9454 & 0.9309 & 0.8873 & 0.8265 & 0.7730 & 0.7418 & 0.7306 & 0.7319 \\ 0.9114 & 0.8996 & 0.8652 & 0.8156 & 0.7662 & 0.7306 & 0.7132 & 0.7106 \\ 0.8832 & 0.8739 & 0.8475 & 0.8085 & 0.7666 & 0.7319 & 0.7106 & 0.7030 \end{pmatrix} \in [0,1] \quad (18)$$

$$AQM_{intra} = \left\lceil \frac{16}{H'(u,v)} \right\rceil = \begin{pmatrix} 16 & 16 & 16 & 16 & 16 & 17 & 18 & 18 \\ 16 & 16 & 16 & 16 & 17 & 17 & 18 & 18 \\ 16 & 16 & 16 & 17 & 18 & 18 & 18 & 19 \\ 16 & 16 & 17 & 18 & 19 & 19 & 20 & 20 \\ 16 & 17 & 18 & 19 & 20 & 21 & 21 & 21 \\ 17 & 17 & 18 & 19 & 21 & 22 & 22 & 22 \\ 18 & 18 & 18 & 20 & 21 & 22 & 22 & 23 \\ 18 & 18 & 19 & 20 & 21 & 22 & 23 & 23 \end{pmatrix} \quad (19)$$

Note that the floating point values in (18) are, indeed, higher than those in the original 2D FWM $H(u,v)$ in (8) for high frequency transform coefficients, while the values for the DC and low frequency AC transform coefficients remain the same. Therefore, these values still represent the visually perceptual importance of transform coefficients in the frequency domain. The integer weighting values in the derived $AQM_{intra}$ in (19) are smaller than those in $QM_{intra}$ in (9). Consequently, lower levels of quantization are applied to video data that is to be decoded and deployed to the 4K VDU.

## 4.0 AQMs in SHVC

The primary objective in SHVC is to allow a video signal to be encoded into a single multilayered bit-stream, in which eight enhancement layers can be embedded using, if required, different spatial resolutions (spatial scalability), different fidelity levels (SNR scalability) and different frame rates (temporal scalability). Although temporal scalability support is available in single-layer HEVC, it has been adopted in SHVC to enable support for combined spatial, and SNR scalability [15].

In comparison with inter-layer prediction in Scalable Video Coding (SVC), JCT-VC has introduced enhancements to inter-layer prediction in SHVC. These advancements are known as inter-layer texture prediction (ILTP) and inter-layer motion prediction. With an emphasis on ILTP, this technique is invoked by including inter-layer reference pictures (ILRP), from the reference layer, in the reference picture lists of the enhancement layer picture [15]. ILTP provides the majority of the coding efficiency improvements in SHVC (in comparison with a two layer simulcast in single-layer HEVC). When inter-layer prediction is enabled, the multiloop coding framework in SHVC requires that reference layers used for inter-layer prediction be fully decoded before the target layer can be decoded [15]. For example, if there are two enhancement layers (ELs) and one base layer (BL), then BL and EL 1 need to be fully decoded so that they can be used as prediction references for decoding EL 2.

In this work, we employ the layers in SHVC, in addition to exploiting inter-layer prediction, to evaluate the efficacy of the proposed AQM method. That is, the AQMs in the ELs are designed for higher display resolutions in comparison with the AQMs in the base layer. Therefore, due to the way in which the AQMs are designed, SNR scalability is provided by the AQMs even if the BL and EL(s) use the same QP. A block diagram, which provides a conceptual overview of AQM's implementation in SHVC, is shown in Fig. 3.

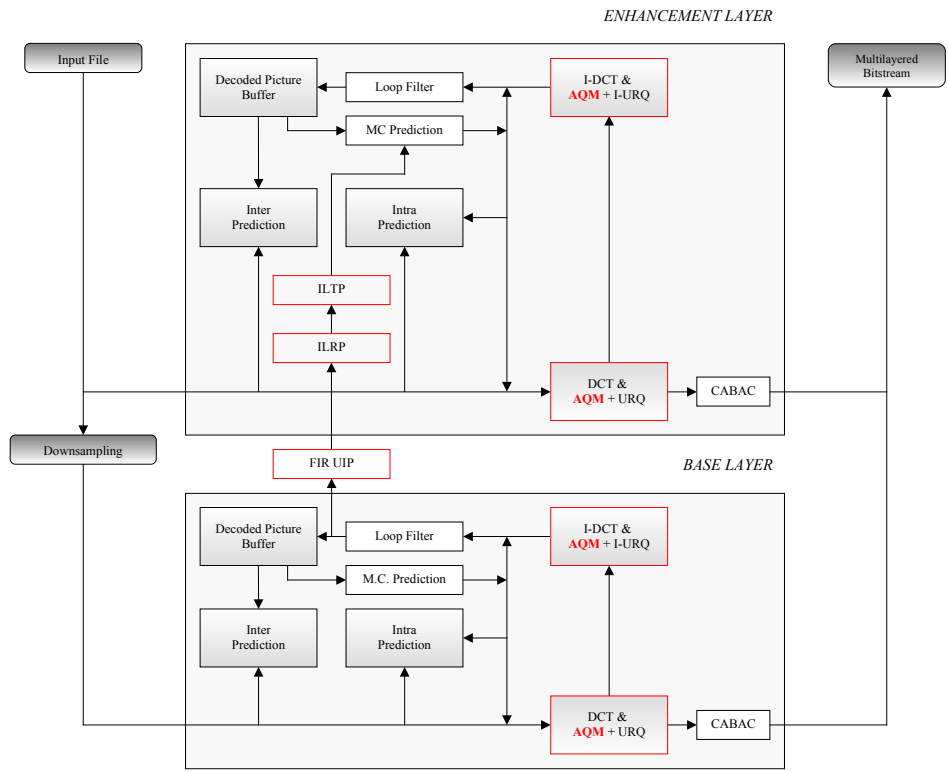

Fig. 3. A block diagram of the AQM technique in SHVC. The AQM technique is highlighted in red. This diagram shows how our method operates on an inter-layer basis in SHVC; it also shows the relationship between the BLs and ELs. Other functions highlighted with a red border, such as the Finite Impulse Response, Upsampling Interpolation Filter (FIR UIF), ILRP and ILTP, are intrinsic features in SHVC designed to exploit inter-layer prediction.

In order to signal the proposed AQMs to the decoder, we take advantage of the way QMs are currently signaled in SHVC. We then employ the upright diagonal scan and a Differential Pulse-Code Modulation (DPCM) encoder to transmit the entries of the 8×8 AQMs for each layer [11, 9]. Note that in SHVC, the scaling list option to enable the default or custom QMs is global. This means that the same QMs are applied to all layers and, therefore, only one set of QMs is signaled to the decoder. We have implemented the AQM technique into SHVC by employing a novel layer specific scaling list option, which allows several sets of default or custom QMs to be signaled to the decoder for multiple layers; i.e., a different set of QMs is signaled for each layer.

## 5.0 Performance Evaluation & Discussion

To evaluate the efficacy of the proposed AQMs, we undertake simulations in SHM 9.0 [11]. In this evaluation, the proposed AQM technique is compared with the default QMs in SHM/HM and a state-of-the-art HVS QM technique developed for HEVC by Sony [5]. The experimental setup is in line with JCT-VC's Common Test Conditions for both SHM and HM [16, 17]. The AQM technique is tested using the All Intra (Main), Low Delay (Main) and Random Access (Main) configurations. The QPs used for the I frames are 22, 27, 32 and 37. The official test sequences used are Traffic (Class A), Cactus (Class B), BasketballDrill (Class C), BasketballPass (Class D) and FourPeople (Class E). Inter-layer prediction is enabled in all tests except for the All Intra (Main) configuration tests on the Class A and Class B sequences. Furthermore, two ELs are used in all tests except for those conducted with Class A and Class B sequences (one EL is used). The AQMs used in the simulations are adapted to the following HD/UHD display resolutions: BL - *1280×720 (HD)*, EL 1 - *3840×2160 (4K)* and EL 2 - *7680×4320 (8K)*.

On the left section in Tables 1, 2 and 3, we tabulate the average BD-Rate performance improvements of the proposed AQM technique compared with the default QMs in SHVC for BL, EL 1 and EL 2, respectively. On the right section in Tables 1, 2 and 3, we tabulate the average BD-Rate improvements of the proposed method compared with the state-of-the-art HVS QM technique, developed for HEVC by Sony, for BL, EL 1 and EL 2, respectively. BD-Rate improvements are calculated as the change in bit-rate when the reconstruction quality, measured by the Peak Signal to Noise Ratio (PSNR) metric, is the same.

Table 1. BL BD-Rate results of the proposed AQM technique compared with anchors. The results in green indicate performance improvements, the results in black indicate no improvements and the results in red indicate BD-Rate inflations.

| Proposed AQM Technique versus Default QMs | | | | | | | | | | Proposed AQM Technique versus Sony QMs | | | | | | | | | |
|---|---|---|---|---|---|---|---|---|---|---|---|---|---|---|---|---|---|---|---|
| Class | All Intra | | | Low Delay B | | | Random Access | | | Class | All Intra | | | Low Delay B | | | Random Access | | |
| | Y % | U % | V % | Y % | U % | V % | Y % | U % | V % | | Y % | U % | V % | Y % | U % | V % | Y % | U % | V % |
| A: BL | -0.6 | -0.5 | -0.3 | -2.3 | -2.6 | -2.6 | -3.0 | -5.0 | -5.4 | A: BL | -2.2 | 0.0 | -0.2 | -3.2 | -3.9 | -4.7 | -4.8 | -6.9 | -7.2 |
| B: BL | -0.4 | -0.1 | 0.0 | -2.1 | -2.2 | -2.5 | -2.9 | -4.5 | -5.0 | B: BL | -1.1 | -1.1 | -1.7 | -2.3 | -4.7 | -6.5 | -3.0 | -6.6 | -8.2 |
| C: BL | -0.4 | 0.2 | 0.2 | -3.1 | -3.3 | -2.9 | -3.6 | -5.4 | -5.0 | C: BL | -2.3 | -3.9 | -4.8 | -1.9 | -5.5 | -5.3 | -4.8 | -11.2 | -10.8 |
| D: BL | -0.4 | 0.2 | 0.2 | -2.1 | -2.3 | -2.6 | -2.3 | -3.9 | -2.9 | D: BL | -2.3 | -0.1 | -3.2 | -1.8 | -2.1 | -3.0 | -2.7 | -5.4 | -5.1 |
| E: BL | -0.2 | 0.2 | 0.3 | -2.9 | -3.6 | -4.4 | -3.3 | -2.8 | -2.8 | E: BL | -0.8 | 0.2 | 0.5 | -3.5 | -4.7 | -5.7 | -4.3 | -3.4 | -3.1 |
| Avg. | -0.4 | 0.0 | 0.1 | -2.5 | -2.8 | -3.0 | -3.0 | -4.3 | -4.2 | Avg. | -1.7 | -1.0 | -1.9 | -2.5 | -4.2 | -5.0 | -3.9 | -6.7 | -6.9 |

Table 2. EL 1 BD-Rate results of the proposed AQM technique compared with anchors. The results in green indicate performance improvements, the results in black indicate no improvements and the results in red indicate BD-Rate inflations.

| Proposed AQM Technique versus Default QMs | | | | | | | | | | Proposed AQM Technique versus Sony QMs | | | | | | | | | |
|---|---|---|---|---|---|---|---|---|---|---|---|---|---|---|---|---|---|---|---|
| Class | All Intra | | | Low Delay B | | | Random Access | | | Class | All Intra | | | Low Delay B | | | Random Access | | |
| | Y % | U % | V % | Y % | U % | V % | Y % | U % | V % | | Y % | U % | V % | Y % | U % | V % | Y % | U % | V % |
| A: EL 1 | -0.8 | -0.5 | -0.1 | -19.7 | -19.1 | -19.0 | -37.4 | -39.7 | -40.1 | A: EL 1 | -2.3 | 0.0% | -0.1 | -52.7 | -56.0 | -56.6 | -56.5 | -58.7 | -59.2 |
| B: EL 1 | -0.4 | -0.1 | 0.0 | -12.3 | -14.2 | -14.4 | -40.4 | -43.7 | -44.5 | B: EL 1 | -1.3 | -1.1 | -1.6 | -36.2 | -42.9 | -43.0 | -50.6 | -55.5 | -56.9 |
| C: EL 1 | 19.0 | 21.1 | 21.5 | -2.9 | -2.6 | -2.2 | -30.0 | -32.0 | -31.9 | C: EL 1 | -45.0 | -47.8 | -48.2 | -19.9 | -24.1 | -23.7 | -44.4 | -49.1 | -49.1 |
| D: EL 1 | -3.2 | -3.7 | -3.6 | -6.3 | -6.9 | -7.0 | -32.5 | -35.2 | -34.5 | D: EL 1 | -52.6 | -56.9 | -57.2 | -28.5 | -31.6 | -31.8 | -39.1 | -42.5 | -42.3 |
| E: EL 1 | -1.6 | -2.0 | -2.0 | -6.0 | -6.7 | -6.9 | -29.3 | -31.3 | -31.1 | E: EL 1 | -33.5 | -35.4 | -35.0 | -6.6 | -6.9 | -7.1 | -33.5 | -35.4 | -35.0 |
| Avg. | 2.60 | 2.96 | 3.16 | -9.44 | -9.90 | -9.90 | -33.92 | -36.38 | -36.42 | Avg. | -26.94 | -28.24 | -28.42 | -28.78 | -32.30 | -32.44 | -44.82 | -48.24 | -48.5 |

Table 3. EL 2 BD-Rate results of the proposed AQM technique compared with anchors. The results in green indicate performance improvements, the results in black indicate no improvements and the results in red indicate BD-Rate inflations.

| Proposed AQM Technique versus Default QMs | | | | | | | | | | Proposed AQM Technique versus Sony QMs | | | | | | | | | |
|---|---|---|---|---|---|---|---|---|---|---|---|---|---|---|---|---|---|---|---|
| Class | All Intra | | | Low Delay B | | | Random Access | | | Class | All Intra | | | Low Delay B | | | Random Access | | |
| | Y % | U % | V % | Y % | U % | V % | Y % | U % | V % | | Y % | U % | V % | Y % | U % | V % | Y % | U % | V % |
| C: EL 2 | 5.5 | 3.8 | 1.5 | -1.1 | -0.7 | -0.7 | -13.8 | -14.0 | -13.8 | C: EL 2 | -18.1 | -19.6 | -19.9 | -9.4 | -10.3 | -10.2 | -34.4 | -36.7 | -36.9 |
| D: EL 2 | 5.2 | 4.6 | 4.7 | -2.6 | -2.3 | -2.3 | -7.6 | -7.9 | -7.8 | D: EL 2 | -15.0 | -17.6 | -17.6 | -11.3 | -11.9 | -12.0 | -17.7 | -18.8 | -18.8 |
| E: EL 2 | -1.6 | -1.8 | -1.7 | -3.8 | -3.9 | -3.9 | -12.5 | -10.3 | -10.6 | E: EL 2 | -20.7 | -20.8 | -20.7 | -4.1 | -4.4 | -4.4 | -20.7 | -20.8 | -20.7 |
| Avg. | 3.03 | 2.20 | 1.50 | -2.50 | -2.30 | -2.30 | -11.30 | -10.73 | -10.73 | Avg. | -17.93 | -19.33 | -19.40 | -8.27 | -8.87 | -8.87 | -24.27 | -25.43 | -25.47 |

As shown in Table 1, in the BL versus BL tests the most significant average coding reductions attained by our method, compared with the default QMs in SHM, are as follows: 3.6% (Y), 5.4% (Cb) and 5.0% (Cr) for the Class C sequence using the Random Access (Main) configuration. In comparison with the QM technique developed by Sony, the most noteworthy average luma and chroma BD-Rate improvements achieved by our method are as follows: 4.8% (Y), 11.2% (Cb) and 10.8% (Cr) using the Random Access (Main) configuration.

In the EL 1 versus EL 1 tests, our technique yielded positive results (see Table 2). Compared with the default QMs in SHM, the largest improvement is recorded for the Class B HD sequence using the Random Access (Main) configuration; the following BD-Rate improvements are achieved: 40.4% (Y), 43.7% (Cb) and 44.5% (Cr). In contrast with the QM technique from Sony, considerable BD-Rate improvements are achieved on the Class A (UHD 4K) sequence using the Random Access (Main) configuration, which are as follows: 56.5% (Y), 58.7% (Cb) and 59.2% (Cr).

The BD-Rate improvements attained for the EL 2 versus EL 2 tests are not as significant as those achieved for the EL 1 versus EL 1 tests (see Table 3). This may be explained by the fact that inter-layer dependencies increase for higher layers [18, 10] and, therefore, the bit-rate and BD-Rate of EL 2 decreases in comparison with the bit-rate and BD-Rate of EL 1 regardless of the QM technique utilized in the layers. In the EL 2 tests, compared with the default QMs in SHM, BD-Rate improvements of 13.8% (Y), 14.0% (Cb) and 13.8% (Cr) are achieved in the Class C, Random Access (Main) simulations. Compared with the QM technique developed by Sony, BD-Rate improvements of 34.4% (Y), 36.7% (Cb) and 36.9% (Cr) are attained in the Class C, Random Access (Main) simulations.

The lower integer weighting values in the proposed AQMs result in lower levels of quantization and, therefore, produce a much higher PSNR value relative to any increases in bit-rate. This is the main reason the proposed method produces high BD-Rate improvements in comparison with the default QMs in HEVC and the Sony QMs. Moreover, the Sony QM technique does not take advantage of the 8×8 QM upsampling and replication process that is intrinsic to the HEVC and SHVC design (see Section 2.0). That is, the Sony QM technique includes the use of 16×16 and 32×32 intra and inter scaling lists.

Our technique is most suited to tests in which the SHVC software is configured for high compression performance. The Random Access configuration represents typical broadcasting and streaming situations, in which bit-streams are expected to enter the decoding process approximately every second [18]. Our technique performs well using the Random Access configuration because of its temporal coding structure. Indeed, the simulation results show that the proposed AQM technique attains the best performance when there is a larger group of B pictures in the GOP structure.

In the All Intra (Main) configuration simulations, compared with the default QMs in SHVC in the BL, our technique yields small BD-Rate gains for the luma component (see Table 1). This is because our method quantizes transform coefficients for intra-predicted residual data using a finer quantization step size, which results in improved PSNR values relative to increases in bit-rate. In comparison with the default QMs in SHVC in the ELs, the proposed method, on average, yields high BD-Rate inflations (see Tables 2-3). Our technique provides more effective quantization on transformed inter-predicted residual values of high resolution video data; i.e., our technique is more suited to the GOP structures of the Random Access and Low Delay inter configurations. Compared with the Sony QM technique, our method attains significant BD-Rate gains in both the BL and ELs. The Sony intra QMs apply much coarser quantization to high frequency components, resulting in decreased reconstruction quality of the finer details in images.

In relation to average encoding and decoding times and compared with the default QMs and the Sony QMs (anchors), the proposed method yields average encoding time reductions of 0.75% and 1.19%, respectively. In addition, our technique yields average decoding time reductions of 4.67% and 2.82%, respectively. Therefore, these simulations demonstrate that our AQM technique does not incur additional computational complexity with respect to encoding and decoding times.

**6.0 Conclusion**

A novel AQM technique for the HEVC standard has been proposed to improve quality reconstruction and reduce the visibility of video compression artifacts on VDUs with high definition display resolutions. In our technique, the integer weighting values in the intra and inter AQMs are adaptive and contingent upon the display resolution of the target VDU. We utilized SHVC SHM 9.0 to evaluate the technique on various sequences of different classes. More specifically, we have created multilayered bit-streams where each layer is aimed at a specific display resolution of a VDU. Compared with anchors, the evaluation reveals that the proposed AQM method yields important coding efficiency improvements for the Random Access (Main) configuration in all tests, with a maximum luma BD-Rate improvement of 56.5%. In terms of encoding and decoding times, our technique yields small improvements in comparison with anchors.